\RequirePackage{fix-cm}
\documentclass{svjour3}                     

\smartqed  

\usepackage{graphicx}

\usepackage[numbers,sort]{natbib}
\usepackage{amsfonts}
\usepackage{amsmath}
\usepackage{amsmath}
\usepackage{subfloat}
\usepackage{subfig}
\usepackage[table]{xcolor}
\usepackage{dcolumn}
\usepackage{multirow}
\usepackage{hyperref}
\hypersetup{colorlinks,
	citecolor=blue
}

\begin{document}


\title{Automatic Modelling of Human Musculoskeletal Ligaments -- Framework Overview and Model Quality Evaluation
}

\author{Noura Hamze\textsuperscript{1} \and 
	Lukas Nocker\textsuperscript{1} \and
	Nikolaus Rauch\textsuperscript{1} \and
	Markus Walzth\"oni\textsuperscript{1} \and
	Fabio Carrillo\textsuperscript{2} \and
	Philipp F{\"u}rnstahl\textsuperscript{2} \and
	Matthias Harders\textsuperscript{1}
}



\institute{
	\textsuperscript{1} Interactive Graphics and Simulation Group, University of Innsbruck, Austria. \\
	\textsuperscript{2} Computer Assisted Research and Development Group, Balgrist University Hospital, University of Zurich, Switzerland.\\
	CONTACT Noura Hamze. Email: nourahamze@gmail.com, Matthias Harders. Email: matthias.harders@uibk.ac.at
}

\date{Received: date / Accepted: date}

\maketitle


\begin{abstract}
	Accurate segmentation of connective soft tissues is still a challenging task, which hinders the generation of corresponding geometric models for bio-mechanical computations. Alternatively, one could predict ligament insertion sites and then approximate the shapes, based on anatomical knowledge and morphological studies. Here, we describe a corresponding integrated framework for the automatic modelling of human musculoskeletal ligaments. We combine statistical shape modelling with geometric algorithms to automatically identify insertion sites, based on which geometric surface and volume meshes are created.
	For demonstrating a clinical use case, the framework has been applied to generate models of the interosseous membrane in the forearm. For the adoption to the forearm anatomy, ligament insertion sites in the statistical model were defined according to anatomical predictions following an approach proposed in prior work. For evaluation we compared the generated sites, as well as the ligament shapes, to data obtained from a cadaveric study, involving five forearms with a total of 15 ligaments. Our framework permitted the creation of 3D models approximating ligaments' shapes with good fidelity. However, we found that the statistical model trained with the state-of-the-art prediction of the insertion sites was not always reliable. Using that model, average mean square errors as well as Hausdorff distances of the meshes increased by more than one order of magnitude, as compared to employing the known insertion locations of the cadaveric study. Using the latter an average mean square error of 0.59$\,mm$ and an average Hausdorff distance of less than 7$\,mm$ resulted, for the complete set of ligaments. 
	In conclusion, the presented approach for generating ligament shapes from insertion points appears to be feasible but the detection of the insertion sites with a SSM is too inaccurate. For this part, another more patient-specific approach should be followed.
\end{abstract}



\section{Introduction}

In orthopedic surgery, an accurate diagnosis and the subsequent preoperative planning of the surgical steps are crucial for the successful restoration of a patient's anatomy. High precision in the preoperative planning and the surgical execution are mandatory for ensuring satisfactory clinical outcomes (e.g.~\cite{furnstahl2010computer,fernandez1982correction}). Three-dimensional, computer-assisted approaches have become the established option as they
contribute to a more comprehensive planning. Providing dynamic biomechanical simulations allows for the analysis of complex pathologies, as well as an evaluation of a larger range of possible intervention strategies. Nevertheless, most preoperative methods are currently limited to simple bone-based models, not taking into account the influence of soft tissue.

For the correct diagnosis of soft tissue influence in case of impairment of the patient function, a simulation framework also integrating the influence of relevant soft tissue structures is required. As an example, the kinematics of forearm pro-supination, and the resulting range of motion, is highly influenced by the interosseous membrane (IOM). The latter is the strongest ligament structure in the forearm, connecting radius and ulna bones~\cite{nagy2008correction,Kitamura2011}. Previous studies have demonstrated that it is the main stabilizing element during forearm motion~\cite{nagy2008correction,Kitamura2011,pfaeffle2000stress,nakamura2000normal}, and thus has to be included in respective simulations. However, creating a geometric simulation model of such soft tissue structures is a challenging task, because of the high inter-subject variation in their shapes and insertion sites. Further, the segmentation of ligaments in medical images is hardly possible in a reliable way, due to low contrast to neighboring anatomy.

Regarding biomechanical simulations, some prior works have  modelled ligaments from patient-specific segmentations. For instance, models of the anterior cruciate ligament of the knee have been generated using morphological operations combined with active contours~\cite{HO2010} or graph cuts with shape constraints~\cite{Lee2014}, applied to MR images. However, processing MR data acquired with clinical protocols is often not possible due to poor resolution and data heterogeneity. Other works have relied on surrogate modeling methods, based on prior knowledge of the ligaments' shapes and locations. A common approach here is to employ single or multiple line segments connecting bones (see e.g.~for the knee~\cite{Bloemker2012,Graf2014}, forearm~\cite{Paen2017}, or 
elbow~\cite{Rahman2018}). In these models, the line segments are considered as linear springs; this is straightforward, but has several limitations, such as the inability to describe complex geometry or to predict nonuniform stresses and strains. As a more accurate alternative, the use of 3D finite element models for simulations has been proposed (e.g.~for analyzing the knee~\cite{Penrose2002,Weiss2005}, ankle~ \cite{Bandak2001}, shoulder~\cite{Zheng2016} or wrist~\cite{SHIMAWAKI2015}). Finally, statistical shape models have also been explored in this context, e.g.~for ACL reconstruction~\cite{Fleute1999} or for the planning of corrective orthopedic surgeries of malunited forearm bones~\cite{Mauler2017}. 

We strongly believe that one reason for the limited clinical translation of functional preoperative planning, which is based on range of motion simulation in the forearm
\cite{kasten2002computer,ward2000role,weinberg2000new,yasutomi2002mechanism,matsuki2010vivo}, is associated with the difficulty of obtaining reliable patient-specific
3D models of the IOM ligaments. Motivated by these limitations, we aimed to develop a
novel framework capable of generating 3D ligament models in an automatic fashion based
on statistical shape models and geometric meshing algorithms.
The work will contribute to improve the accuracy of patient-specific motion
simulations in the forearm. Our approach was evaluated by comparing ligaments segmented from a cadaver study against models of our framework.

\section{Methods}

\subsection{Framework Overview and IOM Test Scenario}

\begin{figure}[h]
	\centering
	\includegraphics[angle=90,width=0.9\linewidth]{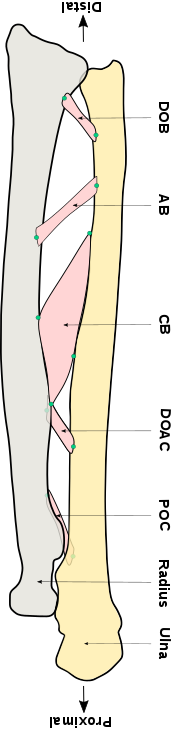}
	\caption{Illustration of the different components of the IOM, with
		corresponding insertion locations indicated as green circles.} 
	
	\label{FIG_iom_components}
\end{figure}

Our ligament modelling framework is comprised of two main parts: a pre-computation phase in which a statistical shape model of the considered anatomy is set up, and a ligament model generation phase, in which ligament meshes for a specific patient anatomy are created. Although the framework is generic, we demonstrate its use in the context of modelling the IOM of the forearm. The latter is the ligamentous complex playing a primary role in maintaining forearm stability during pro-supination motion. Despite its complex anatomy, the IOM shape is often simplified to the most representative component, the so-called central band. However, in some studies \cite{Skahen1997,Fajita1995}, the anatomy is described with more precision; it is classified into several heterogeneous parts, comprising the central band (CB), accessory band (AB), distal oblique accessory cord (DOAC), distal oblique bundle (DOB), and proximal oblique cord (POC). In our work, we also follow this notion. Figure~\ref{FIG_iom_components} indicates the locations of the respective forearm ligaments.

The anatomy of the ligaments can be described also according to their attachment (i.e.~insertion) locations on the bones. We consider these as key anatomical landmarks for our approach. In general, an insertion point is given by the 3D coordinates of a mesh vertex. In Figure~\ref{FIG_iom_components}, also the landmarks are illustrated (as green circles), corresponding to different parts of the IOM. Depending on the ligament width,  a landmark on the bone surface denotes either the center of a thinner band (e.g.~AB), or the start and end location of a wider one (e.g.~CB). In order to label the insertion points, we will below indicate their location as either on the radius (\mbox{\_R}) or on the ulna (\mbox{\_U}); and in case of a wider ligament the respective position on the bone as distal (D) or proximal (P). Thus, the insertion point of the accessory band on the ulna is labeled as AB\_U, while the proximal insertion point on the central band on the radius is specified as CBP\_R. 

Regarding anatomical locations of insertion points, one of the most complete anatomical studies of the IOM was performed by Noda et al.~\cite{Noda2009}, incorporating 30 cadaver forearm specimens. They specified the 3D locations of IOM ligament origin and insertion as distance percentages, given as the ratio of the distance from the distal part of the corresponding bone to its total length. In addition, they provided widths and thicknesses of the structures. This state-of-the-art study will serve as a guide for the construction of the statistical shape model of our framework. Below, in Section~\ref{sec.atlas_creation} we first detail the setup of the statistical model in a pre-processing step. Note that there will be four different submodels, one each for radius and ulna, left and right side. Thereafter, in Section~\ref{sec.ligament_modelling} we outline the use of the shape model to generate new ligament meshes in a patient-specific fashion.

\subsection{Pre-Processing -- Statistical Shape Modelling}
\label{sec.atlas_creation}

\subsubsection{Outline} 

Statistical shape models (SSMs) are a well established approach to describe the variability of a class of similar shapes
(see e.g.~\cite{Heimann:2009:SSM:1538412} for an overview). In SSMs, a key step is to determine variability across a set of constitutive samples, as well as the mean shape. With this information, the SSM permits to approximate any new candidate shape from the same class. 
A key step for SSMs is to establish point-to-point correspondences between input samples. 
To this end, we make use of a Gaussian Process Morphable model (GPMM), as outlined in \cite{Luthi2018}, which extends SSMs and permits incorporating expressive shape priors; in our case landmarks of ligaments on forearm bones. In GPMMs, shape deformations are represented as a Gaussian process, also permitting combination of multiple levels of deformations. Next, we describe the medical datasets from which the SSMs were built.

\subsubsection{Training data}
\label{sec.input_data}

Training data were obtained from CT scans of 18 healthy forearms, acquired using a Somatom Edge Plus scanner (Siemens Medical Systems, Erlangen, Germany); with 1$\,mm$ slice thickness, at 120$\,kV$. 3D meshes of the radii and ulnae bones were segmented via intensity thresholding and region growing (Mimics® Medical, Version 19.0, Materialise 2016, Leuven, 195 Belgium), followed by 3D marching cube reconstruction \cite{Lorensen1987}. The average number of vertices in a bone mesh was 12K, which was further reduced to 3K using mesh decimation methods provided by ParaView \cite{Paraview}. Thereafter, six 3D coordinates for the ligaments' landmarks were manually identified on each bone, guided by Noda's prescribed percentages. The manual labelling was performed by an orthopedic resident at the Balgrist University Hospital Zurich, with the Medical Imaging ToolKit MITK~\cite{MITK}. These details of ligament locations are consistently propagated through the subsequent pipeline.

Next, to be able to create a SSM, consistency has to be ensured across the contributing shapes
by aligning them into the same coordinate system, enforcing the same number of vertices in all meshes, and establishing point-to-point correspondence between the vertices. In a first step, all bone meshes are rigidly transformed, starting with a coarse alignment of all bones according to the main axis of the longest one, followed by a fine alignment via the iterative closest point (ICP) registration \cite{besl1992method}, implemented in the Insight Segmentation and Registration Toolkit (ITK)~\cite{ITK}. For the resulting aligned four sets of input meshes $\cal X$ (one each for radius and ulna, left and right side), we next set up GPMMs to facilitate obtaining the correspondences \cite{Luthi2018}.

\subsubsection{Setup of Gaussian Process Morphable models}
\label{sec.setup_GPMM}
The setup starts with the selection of a reference bone mesh; a typical choice being the average bone shape or one of the aligned input meshes. For each representative bone, denoted by $\Gamma_m \in \cal X$, we assign a Gaussian process  $\mathcal{GP}(\mu, k)$ over the shape domain $\Omega$. Here, function $\mu : \Omega \to \mathbb{R}^3$ encodes the mean deformation (i.e.~shape variation) between instances in $\cal X$; the latter being expressed as the mean square distance between points of two shape surfaces. Further, the Gaussian process kernel $k : \Omega  \times  \Omega \to  \mathbb{R}^{3 \times 3}$ is a covariance matrix function, expressing the spread of deformations inside $\cal X$. Note that a key advantage of using GPMMs is the option to use multi-scale kernel Gaussian processes $k_{ms}$. We employ a smooth kernel $k_g$ for the landmarks as well as a larger scale one $k_h$ for the bones, leading to an expression via a sum of two functions:
\begin{equation*}
	k_{ms}(x, y) = k_g(x,y) + k_h(x,y).
	\label{equ:gau_plus_stat}
\end{equation*}

This permits to conveniently include the ligaments' landmarks with constraint position variations into the model, thus ensuring that these details are included in the shape-prior for the following registration. For setup of the GPMM we employ the open source shape modelling libraries Scalismo~\cite{Scalismo} and Statismo~\cite{Statismo2012}. 
Following~\cite{Luthi2018}, we then make use of the GPMMs for establishing point-to-point correspondence between the training samples in $\cal X$.

\subsubsection{Establishing correspondences}
\label{sec.establishing_correspondance}

The problem of finding correspondences is formulated as a non-rigid registration between a target shape surface $\Gamma_i$ represented by a set of 3D points, and a binary reference image $I_m$, constructed from the Gaussian process reference mesh $\Gamma_m$. In the registration, a transformation function is sought that minimizes the mean squared distance from the reference image to the closest target point in current shape $\Gamma_i$. Applying this transformation to all sample shapes in $\cal X$ yields meshes with the desired properties -- i.e., all meshes have the same number of vertice with established point-to-point correspondences, in the same coordinate system. The whole process is carried out again using ITK implementations. For the optimization in ITK, the Limited Memory Broyden Fletcher Goldfarb Shannon minimization with simple bounds (LBFGSB)~\cite{Byrd1995} is employed. Note that the insertion locations are maintained according to the GPMM formulation. In the final phase, a Principal Component Analysis (PCA)-model is built from the generated intermediate bone meshes with correspondences.

\begin{figure*}[!ht]
	\centering
	\subfloat[PCA-model construction from a set of shapes in correspondence after fitting, landmarks indicated as red circles.]
	{\includegraphics[width=0.45\linewidth]{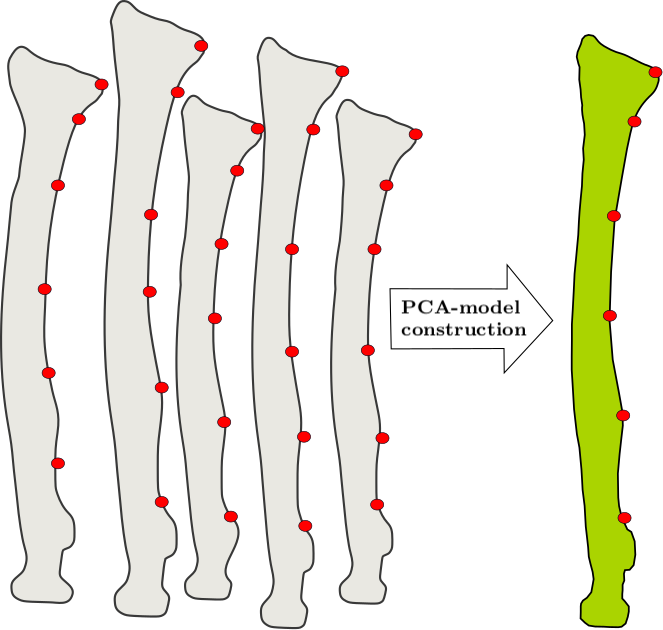}\label{fig_atlas_construction}}
	\hspace{20pt}
	\subfloat[Comparison in 3D of generated landmarks (red) to manually annotated ones (green).]
	{\includegraphics[width=0.40\linewidth]{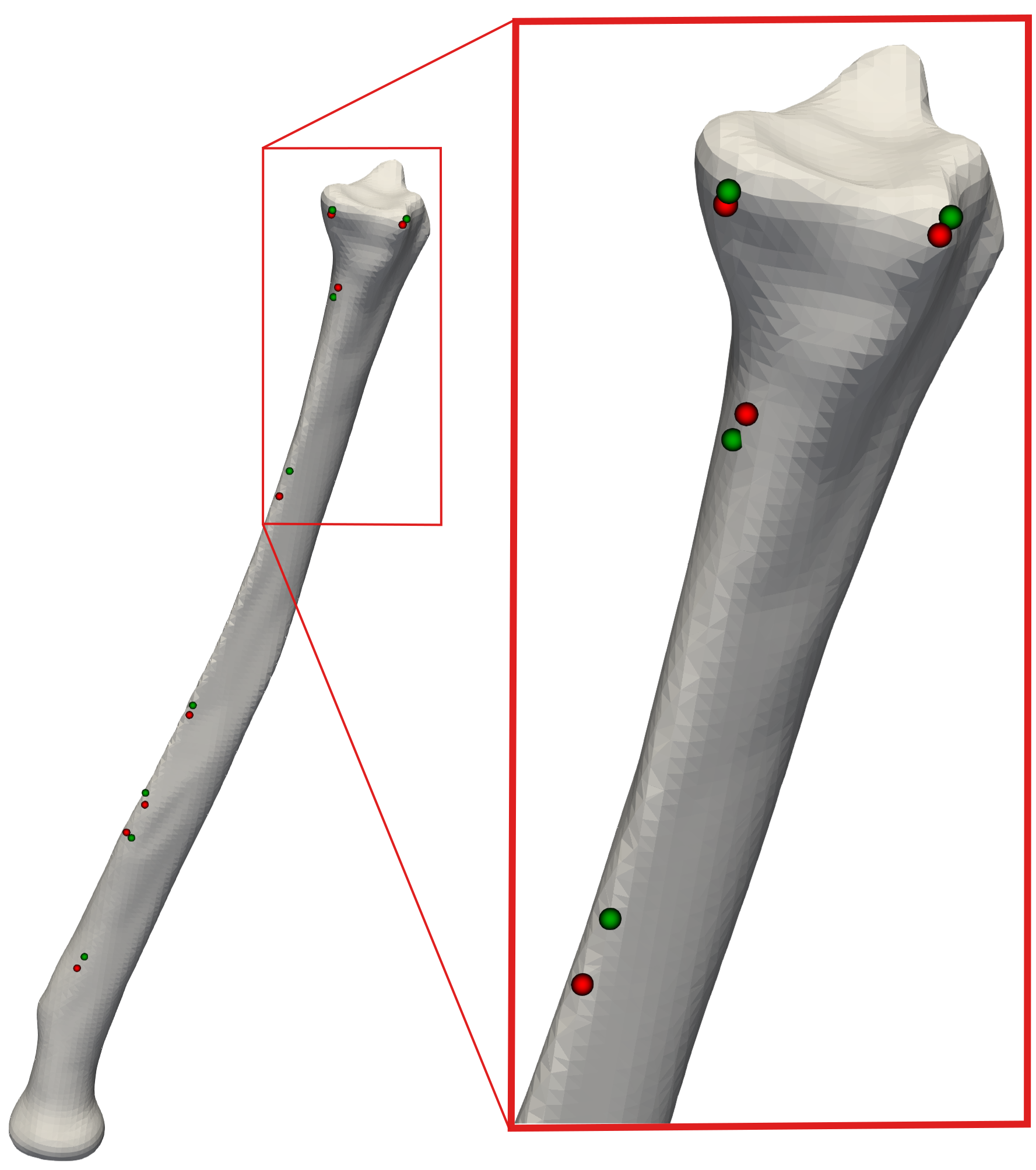}\label{fig_comparison}}
	\caption{Statistical shape model of forearm bones.}  
	\label{shape_model}
\end{figure*}

\subsubsection{Statistical shape model of forearm bones}
\label{sec.SSMs_creation}

Following the PCA scheme, we create the final statistical model. Reference shapes $\Gamma_\mathcal{R}$ with insertion landmarks can be obtained, as linear combination of the mean shape and the most important Eigen-shapes. Note that knowledge about the landmarks' positions is implicitly embedded into the model through a restriction in variation in the intermediate input data. The restriction can be regarded as geometric constraints fixing parts of the bone at the ligament connection sites. Figure~\ref{fig_atlas_construction} illustrates the PCA-model resulting from a set of meshes in correspondence with landmarks. In order to assess the accuracy of the shape model,  we carried out a cross-validation study with leave-one-out testing. We found a mean error of 3$\,mm$ in the landmarks' locations across all data samples. Figure~\ref{fig_comparison} also qualitatively compares landmarks obtained with our framework with manually annotated ones.

\subsection{Patient-Specific Ligament Model Generation}
\label{sec.ligament_modelling}

\subsubsection{Outline}

\begin{figure*}[!ht]
	\centering
	\subfloat[Alignment of target $\Gamma_\mathcal{T}$ (yellow; shown before and after) to reference $\Gamma_\mathcal{R}$ (green).]
	{\includegraphics[height=45mm]{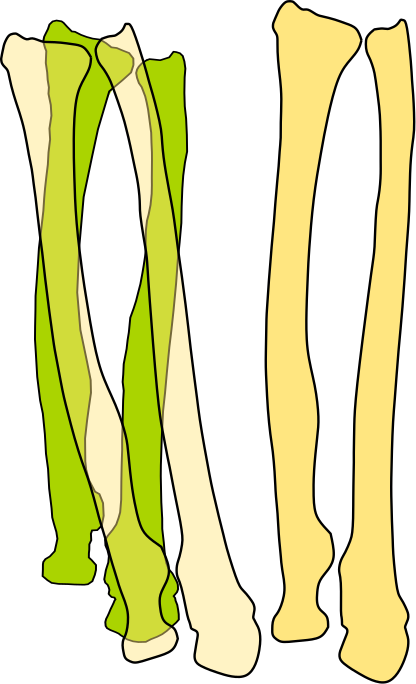}\label{fig_pipeline_alignment}}
	\hfill
	\subfloat[Fitting process via establishing correspondences.]
	{\includegraphics[height=45mm]{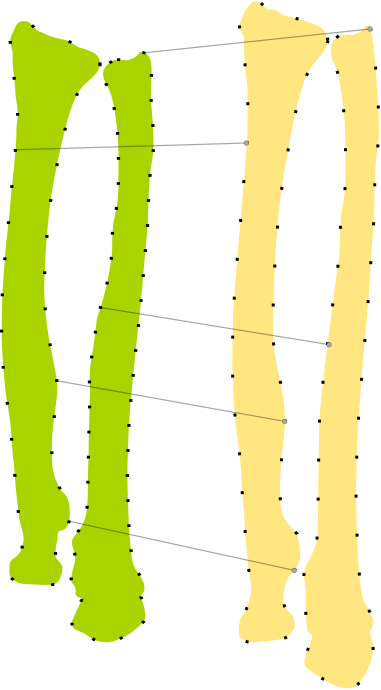}\label{fig_pipeline_fitting}}
	\hfill
	\subfloat[Projection of landmarks.]
	{\includegraphics[height=45mm]{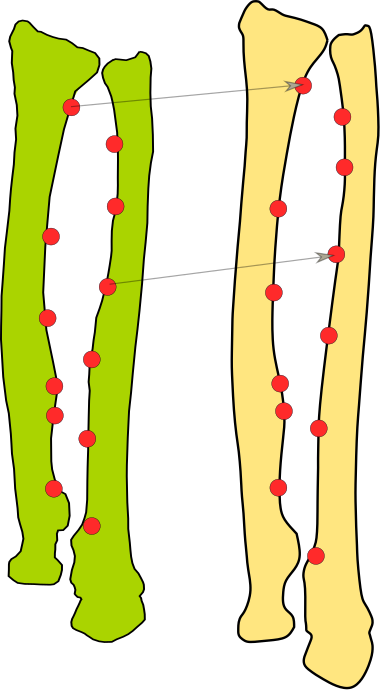}\label{fig_pipeling_projection}}
	\hfill
	\subfloat[Creation of ligament meshes (with an instance shown for two insertion points).]
	{\includegraphics[height=45mm]{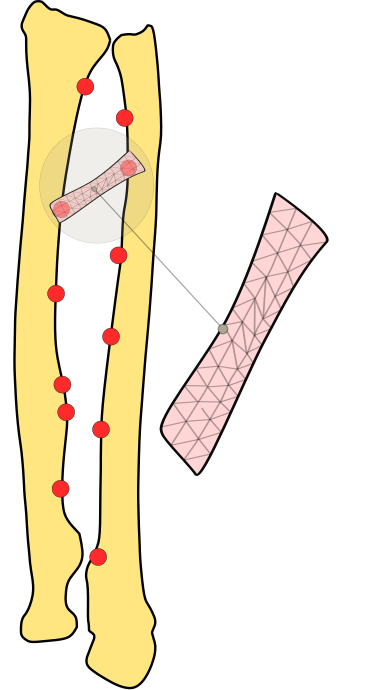}\label{fig_pipeline_modelling}}
	\caption{Overview of the modelling pipeline.}  
	\label{fig.overview}
\end{figure*}

The target of the model generation process is the automatic creation of ligament meshes, based on the known bone geometry of a specific patient, but without any knowledge about the ligaments. To predict the ligament insertion sites, we rely on the SSMs described in Section~\ref{sec.SSMs_creation}. In the first step, landmarks of ligament insertions are automatically transferred to the 3D patient-specific bone meshes which can be obtained by segmentation of medical image data. In the second step, 3D surface models represented by triangles or, if needed, volumetric meshes represented by tetrahedra are generated, approximating the shapes of the ligaments. The modelling is based on the identified landmarks as well as on additional anatomical parameters. The key steps in the modelling pipeline are overviewed in Figure~\ref{fig.overview}.

\subsubsection{Landmark transfer}
\label{sec.landmark_transfer}

In order to transfer a set of landmarks $\mathcal{P}$ to a patient target bone shape $\Gamma_\mathcal{T}$, the latter is first aligned to the reference shape $\Gamma_\mathcal{R}$ of each bone-specific SSM, via the ICP algorithm. Next, the target shape is fitted to $\Gamma_\mathcal{R}$, following the same process used before for establishing point-to-point correspondences (see Section~\ref{sec.establishing_correspondance}). Finally, the set of landmarks found on the SSM are projected onto the target patient bone shape. For each landmark in the reference shape the closest vertex on the aligned and fitted target shape $\Gamma_\mathcal{T}$ is found and labeled as that respective ligament landmark. The output of this phase is a patient-specific bone mesh with all ligament insertion points.


\subsubsection{Ligament shape modelling}
\label{sec.surface_modeler}

3D meshes of ligaments are generated based on the patient-specific bone meshes and associated landmarks of the previous step. As additional information, thickness of ligaments, and potentially also their width has to be supplied. The process is again outlined for the test scenario of the IOM, but transfers directly to any other anatomy. 

\begin{figure*}[!ht]
	\centering
	\includegraphics[width=0.35\linewidth,angle=11]{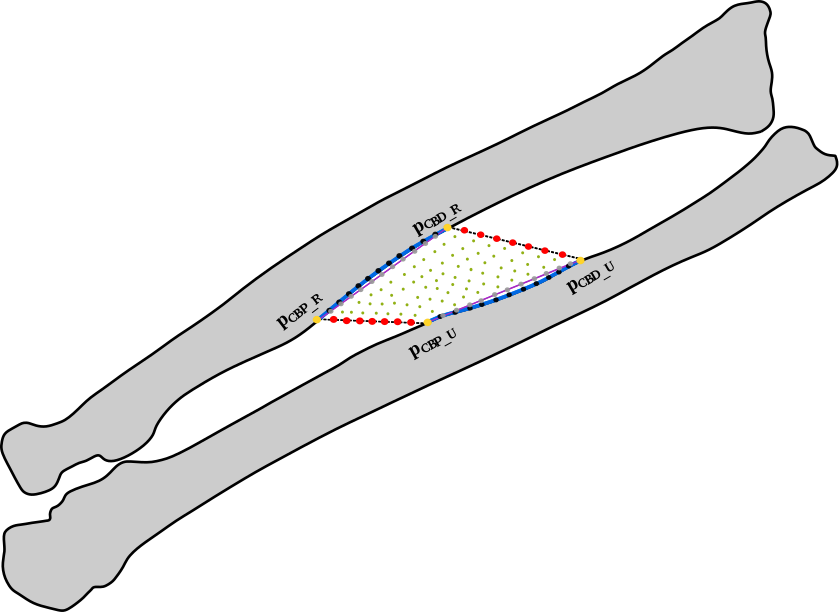}
	\includegraphics[width=0.6\linewidth]{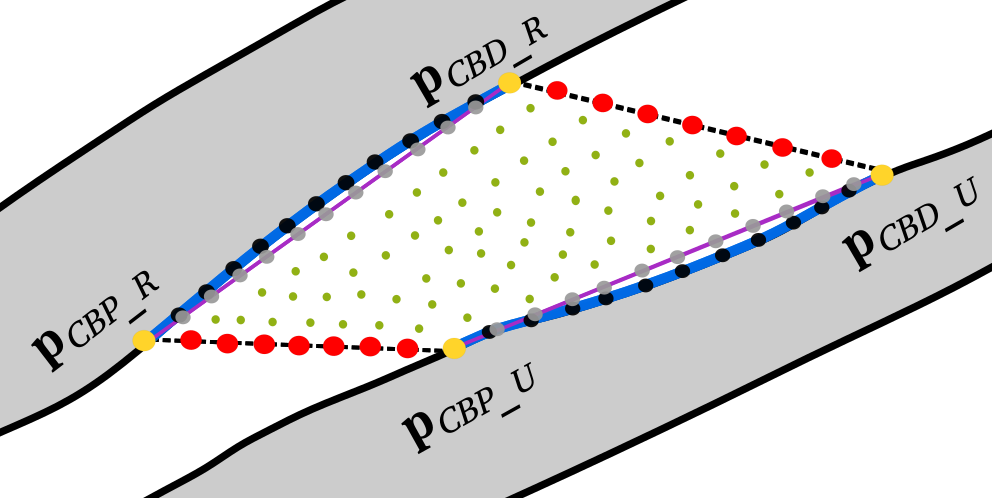}
	\caption{Vertices generated for building the initial ligament surface triangle mesh (in order of creation; \textit{orange}: ligament endpoints, \textit{red}: sampled traversal segments, \textit{grey}: sampled bone-aligned segments, \textit{black}: projected samples, \textit{green}: sampled ligament sheet).}
	\label{fig_vertices}
\end{figure*}

The process is carried out in several substeps. First, four vertices are identified -- the proximal and distal ends of a ligament on each bone. Next, pairs of these are connected, yielding four line segments forming a quadrilateral surface. The line segments are then regularly discretized, yielding a set of vertices on the quadrilateral boundary; some vertices are further projected onto bones surfaces for a better fit. Based on all boundary vertices, new vertices are then generated within the quadrilateral. A triangle mesh is finally generated from all vertices and perpendicularly extruded. Figure~\ref{fig_vertices} indicates the vertices generated in this process, in case of the CB. In the following, the individual shape modelling substeps will be addressed with more detail.

The objective is to model a 3D surface of a ligament connecting two bone surfaces.
As mentioned above, the starting points are four vertices located at the proximal and distal ends of a ligament on the radius and ulna meshes (orange vertices). According to the setup of the SSMs, in case of the CB these vertices are directly given by the four landmarks; i.e.~vertices
$\textbf{p}_{\scalebox{.7}{\textit{CBP\_R}}}$, 
$\textbf{p}_{\scalebox{.7}{\textit{CBD\_R}}}$,
$\textbf{p}_{\scalebox{.7}{\textit{CBP\_U}}}$, 
$\textbf{p}_{\scalebox{.7}{\textit{CBD\_U}}}$, 
which in addition also yield the widths of a ligament along a bone directly on the surface. For the remaining ligaments (DOB, AB, DOAC, POC), only a mid-point landmark is available from the SSMs. In this case the four vertices at the ligament ends are found via the  bone main axes direction vectors 
$\textbf{d}_{\scalebox{.7}{\textit{R}}}$ or 
$\textbf{d}_{\scalebox{.7}{\textit{U}}}$ (pointing from proximal to distal end), and associated ligament widths per bone $w_{\scalebox{.7}{\textit{R}}}$ or $w_{\scalebox{.7}{\textit{U}}}$. The former can be computed from the bone meshes, the latter have to be user-defined. This also results in four vertices; for instance, for AB we determine as one vertex 
$\textbf{p}_ {\scalebox{.7}{\textit{{ABD\_R}}}} = 
\textbf{p}_ {\scalebox{.7}{\textit{{AB\_R}}}} + 0.5\cdot 
w_{\scalebox{.7}{\textit{R}}} \cdot 
\textbf{d}_{\scalebox{.7}{\textit{R}}}$.

Next, forming pairs of some of the four vertices -- i.e.~on each bone (magenta line segments), as well as traversally between the proximal and distal ends (stippled black line segments) -- yields a quadrilateral surface. Note that the former line segments may intersect the bone meshes or the endpoints may not directly touch the bone surfaces. Subsequently, the line segments are regularly subdivided yielding additional vertices. The sampling is controlled via user-supplied parameters, indicating the intra-(along the bone) and inter-bone number of samples per ligament; in our experiments intra- and inter-bone sampling were, for instance in the case of CB, set to 10 and 12, respectively. The resulting 3D vertices are depicted as grey and red circles on the line segments. 

As previously indicated, the line segment along each bone (magenta), and thus the sampled vertices (grey), will likely not be located directly on the bone mesh. Thus, in an additional step, these vertices are first projected onto the nearest bone mesh triangle along its normal whereafter the nearest existing vertex on the bone surface is finally found (black vertices). The reason for using existing vertices is the easier handling of boundary conditions in a potential subsequent biomechanical simulation, e.g.~using finite element meshes. Overall, this results in a piecewise linear approximation of the ligament contact on the bone surface (blue).

\begin{figure*}[!ht]
	\centering
	\subfloat[Generated intermediate triangle mesh.] {\includegraphics[width=0.4\linewidth]{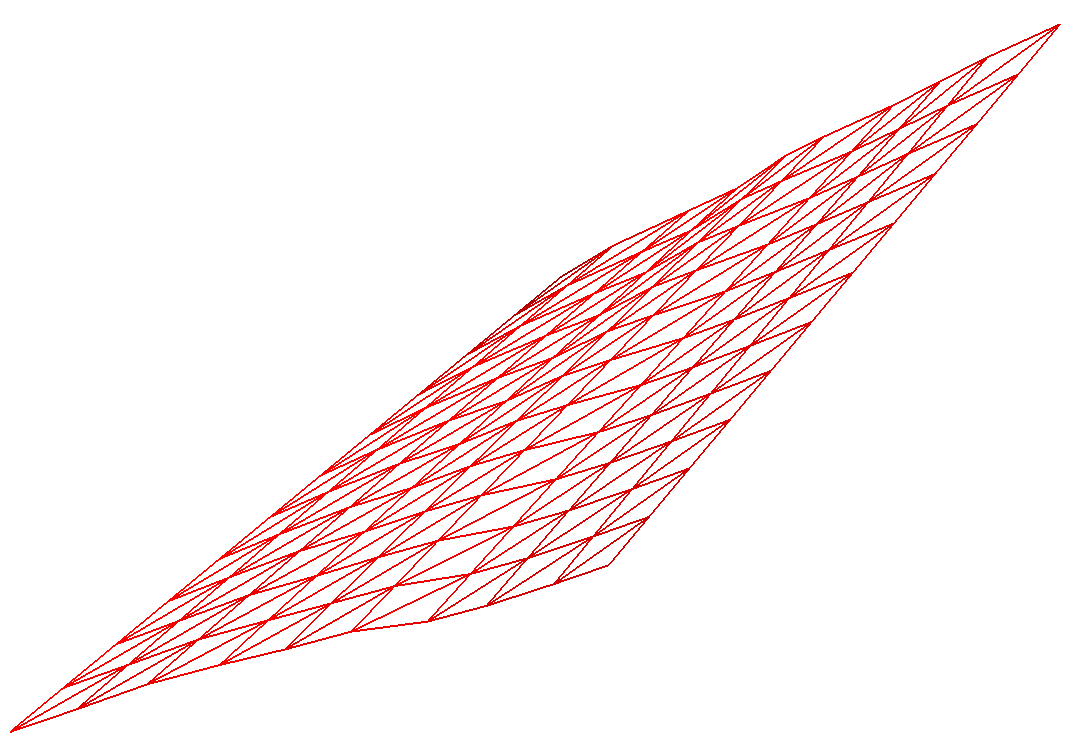}\label{fig_CB_modeling_a}}
	\hfill
	\subfloat[Visualizations of tetrahedral 3D volume mesh created in two steps.]
	{\includegraphics[width=0.55\linewidth]{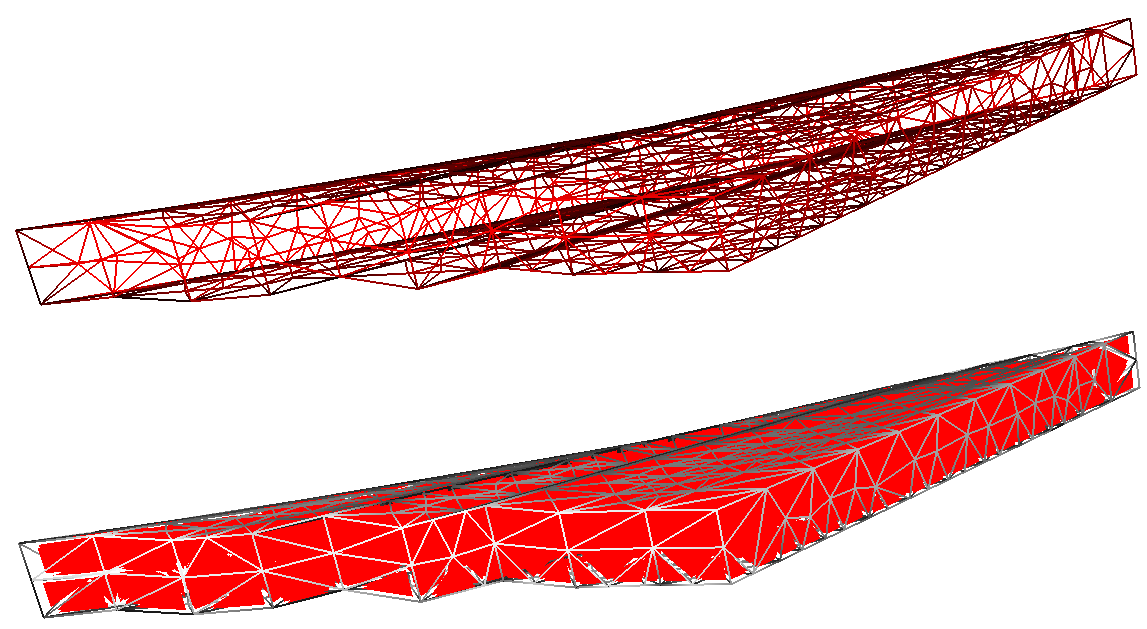}\label{fig_CB_modeling_b}}
	\caption{Examples of generated ligament 3D surface and volume models (in this case, for the CB of dataset ${\cal DS}_3$).}  
	\label{fig:meshes}
\end{figure*}

Based on the vertices -- i.e.~those on the transversal segments (red) and the projected ones (black) -- additional 3D vertices are then generated at regular locations (green), via linear interpolation of positions. Next, using the incremental builder functionality of the Computational Geometry Algorithms library CGAL~\cite{CGAL}, a polygonal (triangle) surface is constructed from all vertices. Note that we employ a halfedge data structure to maintain incidence information of vertices, edges, and faces, using the CGAL Halfedge package \cite{cgalP}. An example of such a resulting triangle mesh is illustrated in Figure~\ref{fig_CB_modeling_a}.
Finally, to incorporate the ligament thickness, the mesh is extruded into normal direction, via the Linear Extrusion filter of ParaView \cite{Paraview}. A parameter for the thickness also has to be supplied; in our experiments we employed values in the range of 1 to 4 $\,mm$. The overall result is a 3D piecewise linear mesh, approximating the surface of a single ligament. 

As an additional step, the surface triangle mesh of a ligament can be further transformed into a volumetric tetrahedral mesh. 
This is carried out with the Constrained Delaunay-based quality tetrahedral mesh generator TetGen~\cite{TetGen}.
In TetGen, the mesh quality can be controlled either by enforcing a maximum volume or a minimum radius value for each element. As an example, we have used radii in a range of 0.6 to 1 $\,mm$. \textit{Steiner} points are added inside the ligament mesh volume to satisfy the imposed constraint. An example mesh is visualized in Figure~\ref{fig_CB_modeling_b}. 

\section{Experiments and Results}

\subsection{Accuracy of Automatically Generated Landmarks}
\label{sec.first_eval}

In a first study, we examined the accuracy of the landmarks that were automatically generated with our presented framework. To do so, we obtained ground truth data via a cadaveric study (see also \cite{Carrillo2019}). In the latter, radial and ulnar insertion locations of the ligaments were visually identified on five forearm specimens (two female, three male); associated datasets are denoted below as ${\cal DS}_1$ to ${\cal DS}_5$. The individual ligaments of the IOM were marked by a surgeon with titanium ligation clips, via a clip applicator tool (Ethicon endo-surgery, LLC, USA). Note that the POC could not be labeled in any of the samples. Moreover, in the specimens either the DOAC or the DOB was present, but not both; the former was found in four cases, the latter only in one. This finding is in line with current anatomical knowledge about variation in forearm ligaments (see e.g.~\cite{Kitamura2011,Noda2009}).

\begin{figure}[ht]
	\centering
	\includegraphics[width=0.8\linewidth]{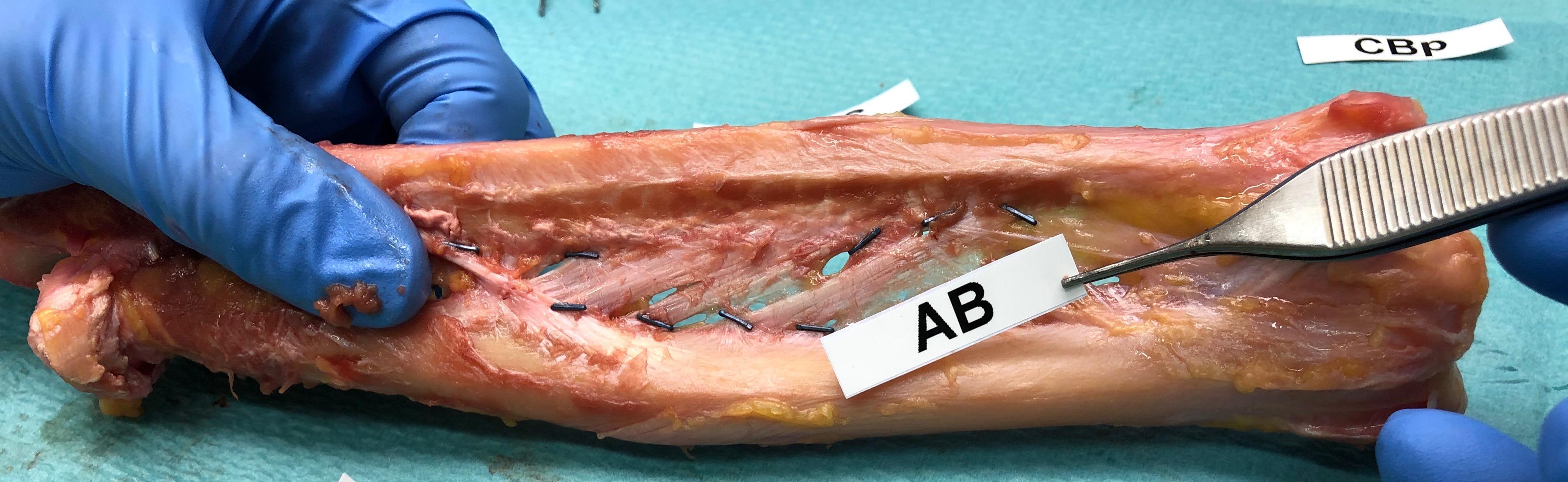}
	\caption{Annotation of ligaments with metallic clips in cadaver forearms; ${\cal DS}_1$ shown as example.}
	\label{Fig.specimen}
\end{figure}

For annotation of DOB, AB, and CB four clips were employed, at the proximal and distal insertion locations of the ligament attachments on both bones, respectively. Further, due to its narrow width, the DOAC was annotated using only one metal clip per bone. The markers were placed at the centers of the attachments of its main fiber bundle.
An example image of an annotated specimen is shown in Figure~\ref{Fig.specimen}.
Thereafter, Micro-CT as well as CT images were obtained from the annotated cadaveric arms. In the resulting image data, all IOM insertion locations were then manually segmented; the latter subsequently serving as ground truth for the  comparison.

Next, using our proposed framework, we automatically obtained an additional set of landmarks via our SSM. As outlined, the input to the process was the bone surface meshes created from segmentations of the CT scans of the five cadaver arms. The output was the estimated landmarks of the IOM ligaments, matching the input geometries. 

\begin{figure*}[!ht]
	\centering
	\subfloat[Mean square error in landmark mid-point locations in $mm$, comparing SSMs landmarks with those from cadaver study.] {\includegraphics[width=0.5\linewidth]{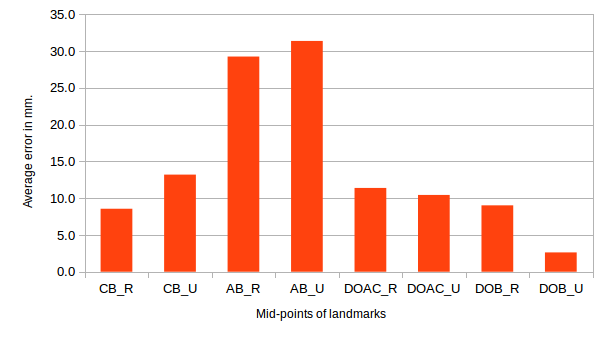}\label{fig_error_in_landmarks_a}}
	\subfloat[Error in widths for CB in $mm$.] {\includegraphics[width=0.5\linewidth]{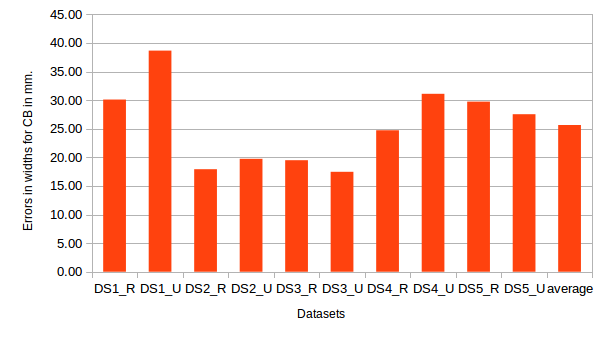}\label{fig_error_in_widths_b}}
	\caption{Quantitative evaluation of landmarks locations and ligament width. }  
	\label{Fig.GT_eval1_diagrams}
\end{figure*}

For further processing, in both the SSM-predicted and ground truth datasets, we computed for the wider ligaments (i.e.~those with four marked locations; two proximal and two distal) the average mid-points. Based on this, we finally determined the $\ell^2$-vector norm of the distances between the locations in the ground truth data and in the framework output. The averages of the distance errors are visualized per insertion location mid-point in millimeters in Figure~\ref{fig_error_in_landmarks_a}. Note that due to the presence or absence of ligaments, some values are only based on a single measurement (i.e.~the DOB).

Excluding the AB, the average errors range between 2.5$\,mm$ and 12.0$\,mm$; i.e.~for CB, DOAC, and DOB. In contrast, the AB exhibits considerably larger errors of about 30$\,mm$. This large value is mainly caused by dataset ${\cal DS}_3$, for which the respective ligament was exceptionally proximal. 

Besides the insertion locations, the accuracy of the generated models can be assessed by examining the widths of the ligaments. 
Figure~\ref{fig_error_in_widths_b} shows exemplary the width of the CB for the five specimens, separately for radius and ulna. An average width of about 
35$\,mm$ was found for this band in the cadaver dataset. Note that this differs considerably from the 9.7$\,mm$ average reported by Noda et al.


\subsection{Similarity of Generated Ligament Meshes}
\label{sec.second_eval}

In the second experiment, we examined the geometric similarity of ligament surface triangle meshes, generated following different strategies. As ground truth we employed meshes, which were manually segmented by a medical student in the previously described cadaveric study. Starting points for these were raw ligament meshes which were then cut employing ParaView~\cite{Paraview} into three sub-components, according to the provided metallic clip annotations (see Section~\ref{sec.first_eval}) on both radius and ulna. Note that when only the ligament mid-points were known, then widths were set according to quantitative measurements of the cadaver study. The resulting meshes were further improved, e.g.~by filling holes. Overall, this resulted in three triangle surface meshes, for the respective ligaments present in the forearm specimen datasets ${\cal DS}_1$ to ${\cal DS}_5$.

\begin{figure}[ht]
	\centering
	\subfloat[Modelling with landmarks from our statistical model.]
	{\includegraphics[width=\linewidth]{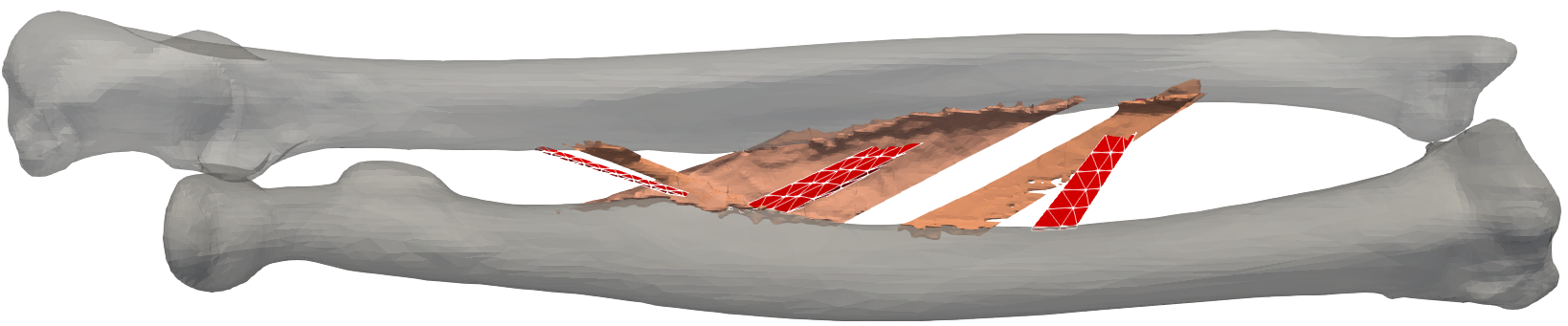}}\\
	\subfloat[Modelling with landmarks given via clip annotation.]
	{\includegraphics[width=\linewidth]{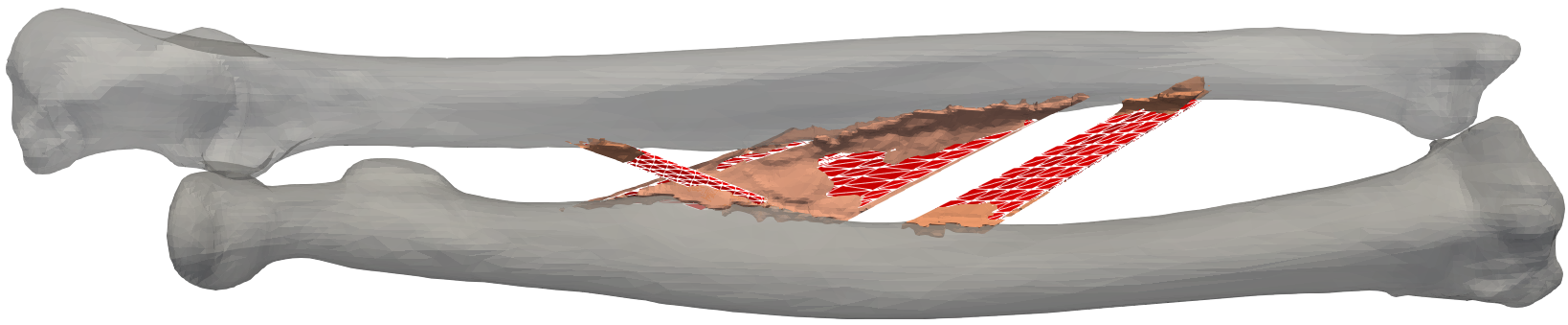}}
	\caption{Comparison of modelled ligament meshes (red) against ground truth (beige) for example dataset ${\cal DS}_1$ (in this case comprising AB, CB, DOAC). }
	\label{Fig.GT_eval1}
\end{figure}

To this ground truth ${\cal M}_{GT}$ we then compared two sets of ligament meshes generated with our geometric modelling framework. For the first, we obtain insertion locations with our newly developed statistical framework. 
The widths of the ligaments were given according to the findings by Noda et al.~\cite{Noda2009}; i.e.~9.7$\,mm$ for CB, 3.2$\,mm$ for DOAC, and 4.4$\,mm$ for DOB, respectively. Further, in the case of AB, an experimental value of 7$\,mm$ was assigned, since this measurement was not provided in their study. Similarly, thickness was set according either to their data, or to the values of our cadaver study.
For the second set of meshes, the generation process was repeated, however, this time using the ground truth locations of the metallic clips as insertion points. Moreover, widths and thicknesses were set as measured in \cite{Carrillo2019}. Overall, this resulted in two additional sets of meshes ${\cal M}_{sta}$ and ${\cal M}_{clp}$, which were compared against the ground truth. To illustrate the resulting  triangle mesh surfaces, Figure~\ref{Fig.GT_eval1} depicts these exemplary for dataset ${\cal DS}_1$. Meshes are rendered for the ligaments DOAC, CB, and AB, showing the ground truth (beige) against the modelled ligament meshes (red). Ligaments generated with insertions from the statistical framework are shown on top, those using the annotated clips instead on the bottom. Note that in the latter case the visual similarity is higher, since the same insertion locations and widths were employed.

In order to quantify the similarity between a modelled triangle surface mesh ($M$) and the ground truth ($GT$), we computed four standard shape similarity metrics:
\begin{enumerate}
\item the \textit{mean Euclidean} distance $d_{ME}$ -- for all vertices in $M$ to their nearest neighbor vertices in $GT$; 
\item the \textit{average symmetric surface} distance $d_{AS}$ -- averaging all distances, from vertices in $M$ to nearest neighbors in $GT$ and vice versa; 
\item the \textit{root mean square} distance $d_{RMS}$ -- given as square root of the average of all square distances from vertices in $M$ to nearest neighbors in $GT$ and vice versa;
\item the common \textit{Hausdorff} distance $d_{HD}$ between two points sets (see e.g.~\cite{Rote1991}).
\end{enumerate}
The latter is given as the largest of all distances from a vertex in one mesh to the closest point in the other mesh. Applying these metrics, we obtain similarity measurements for both sets of modelled ligaments, in comparison to the ground truth. The data are compiled in Table~\ref{table_shape_similarty_evaluations}. For each of the five forearm specimens the similarity metrics are provided (in $mm$) per ligament and modelling approach. Also, the average over all values is specified.

As can be seen, surface meshes modelled using statistical prediction of insertion locations (${\cal M}_{sta}$) exhibit larger errors compared to using the exact insertion locations (${\cal M}_{clp}$). This is expected, since in the latter case the error to the ground truth is only due to differences in shape of the surfaces. In contrast, 
in the former case it is also resulting from differences in the ligament attachments and widths. Clear outliers (in bold) are the $d_{ME}$ metrics for the AB in ${\cal DS}_3$ and ${\cal DS}_4$. The main reasons is the identification of the AB ligaments in those cadaver specimens on the proximal instead of the distal portion of the bones. Such a case was not included in the SSM training data. This underlines the high morphological variation in the forearm ligament anatomy. The relatively low errors for ${\cal M}_{clp}$ indicate that the ligament mesh modelling process can provide reasonable results, given insertion locations and widths of sufficient accuracy.

\begin{table}[ht]
	\begin{center}
		\caption{Similarity errors in $\,mm$ between ground truth and ligament meshes modelled with our pipeline, assuming different insertion locations and widths; compiled for three ligaments in five cadaver specimens (note DOAC vs.~DOB); outliers are shown in bold.}
		\label{table_shape_similarty_evaluations}
		\vspace{2mm}
		\begin{tabular}{|c|c||cc|cc|cc|cc|}
			\hline
			\rowcolor{gray!25} Data- & 
			Liga- & 
			\multicolumn{2}{c|}{$d_{HD}$} & 
			\multicolumn{2}{c|}{$d_{ME}$} & 
			\multicolumn{2}{c|}{$d_{AS}$} &  
			\multicolumn{2}{c|}{$d_{RMS}$} \\ 
			\cline{3-10} 
			\rowcolor{gray!25} set &  ment  & 
			${\cal M}_{sta}$ & ${\cal M}_{clp}$ & 
			${\cal M}_{sta}$ & ${\cal M}_{clp}$ & 
			${\cal M}_{sta}$ & ${\cal M}_{clp}$ & 
			${\cal M}_{sta}$ & ${\cal M}_{clp}$ \\ 
			\hline
			\multirow{3}{*}{${\cal DS}_1$} 
			& {\small AB}   &  22.14 & 5.74 & 8.97 & 0.24 &5.92 &  0.73 & 7.875 & 1.161 \\
			& {\small CB}   &  31.64 & 7.48 & 0.33 & 0.28 &5.45 & 0.99  & 7.73& 1.38\\
			& {\small DOAC} &  13.40 & 7.20 &  4.55 & 0.49 &  3.11 & 1.50 & 3.87& 1.93\\
			
			\hline 				
			\multirow{3}{*}{${\cal DS}_2$} 
			& {\small AB}   & 24.90 &7.70 & 2.91 & 0.38 &   7.01 & 1.45 & 8.85 &2.23\\
			& {\small CB}   & 13.82 & 9.61 &  0.62 & 0.84 &  4.06 & 1.74 & 4.81& 2.37\\
			& {\small DOAC} & 20.17 &3.83 &  2.89 &0.65 &   5.00 & 1.37 & 6.91& 1.65\\
			
			\hline 				
			\multirow{3}{*}{${\cal DS}_3$} 
			& {\small AB}   & 95.98 & 3.13 & \textbf{3067.31} & 0.65 &    70.83 & 0.98 & 71.8& 1.2\\
			& {\small CB}   & 13.67 & 5.73 &  0.62 & 0.45 &  2.30   & 0.77 & 3.29& 1.03\\
			& {\small DOAC} & 20.28 & 7.22 &  99.98 & 0.47 &  11.31   & 1.51 & 11.56&2.02\\
			
			\hline 				
			\multirow{3}{*}{${\cal DS}_4$} 
			& {\small AB}   & 75.41 & 2.23 & \textbf{2873.54}  & 0.37 & 57.06 & 0.63 & 57.69 & 0.76 \\
			& {\small CB}   & 33.51 & 8.02 &  0.84  & 0.61 &   6.86  & 0.99 & 9.67 & 1.32\\
			& {\small DOAC} & 4.19 & 6.57 & 1.12    & 0.81 &  1.62 & 1.25 & 2.02&1.63\\
			
			\hline 				
			\multirow{3}{*}{${\cal DS}_5$} 
			& {\small AB}   & 46.22 & 9.47 & 59.77 & 0.60 &  19.53 & 1.48 & 22.12& 2.04\\
			& {\small CB}   & 27.71 & 6.81 &  13.75 & 0.75 & 5.18  & 1.19 & 7.54&\ 1.45\\
			& \textbf{\small DOB} & 15.95 & 13.75 & 3.37 & 1.25  & 4.82 & 2.19 & 5.76& 3.1\\
			
			\hline
			\cline{3-10}
			\multicolumn{2}{|c|}{\textit{Average}} & 30.60  & 6.97 & \textbf{408.82 }& 0.59 & 14.00 & 1.25 & 15.44 & 1.69 \\	
			\hline 				
		\end{tabular}
	\end{center}
\end{table}

\section{Discussion \& Conclusion}

We have outlined an automatic framework for patient-specific musculoskeletal ligament modelling. To this end, we combined statistical shape models with landmark transfer and mesh generation algorithms. As a proof of concept we applied the generic framework to the test scenario of the interosseous membrane of the forearm. For this, a statistical shape model was created based on forearm image data of 18 healthy subjects. Our framework was then evaluated using an additional dataset of five annotated cadaver forearm specimens. Ligament insertion locations and shapes based on our framework have been compared to the reference specimens.

The landmark transfer algorithm is capable of automatically generating landmarks for any ligament based on a statistical anatomical shape model. Regarding the latter, for our test scenario the training dataset was still relatively small, with only 18 samples. Nevertheless, increasing the number of samples on the same order of magnitude may not directly lead to a large accuracy improvement, due to the high anatomical variation of the IOM ligaments and insertion locations among humans. Our study indicates that using distance ratios on average bone shapes, as proposed in previous work such as Noda et al.~may not be of sufficient precision for a reliable clinical application. Another source of inaccuracy may arise from the labeling process, both for the SSM training samples as well as the cadaveric study; this step was performed by only a single person, which may also introduce deviations from the exact locations. Further, the number of samples in the ex-vivo experiment was low and may have to be increased to allow for an improved quantitative assessment. Finally, the separation of the segmented IOM complex into separate bands likely also has suffered from inaccuracies.

As outlined in our study, errors were still found between ligaments modelled according to SSM generated insertion points and corresponding segmentations of ground truth data. Nevertheless, given accurate ligament attachments and widths, our modelling framework can generate shapes with reasonable accuracy. Still, smaller error is introduced according to the anatomical distribution of thickness and width of the ligaments, which are irregular across axial and medial direction \cite{Carrillo2019,mcginley2001mechanics}. Another source of possible inaccuracy in our modelling algorithm can be the ligament-bone connection boundaries. These were defined by line segments along the bone surface, between two given vertices. Modelling these as spline curves instead may increase accuracy. 

Nevertheless, to the best of our knowledge, our study is the first to provide quantitative comparisons of automatically modelled interosseous membrane ligament shapes with meshes based on ex-vivo data. Moreover, we 
have presented a framework for automatic ligament mesh generation; it is generic and can easily be applied to the modelling of other ligaments in the human body. It remains to be seen, which effect errors in the geometric shapes of ligaments will have on the overall forearm motion behavior in biomechanical simulations. 

Overall, the described development is part of a larger research project focusing on improving the quality of preoperative planning in upper extremities. The introduced framework constitutes one step towards a more accurate representation of ligaments for simulations of forearm motion. The inclusion of soft tissues into the preoperative planning is generally a challenging task, which requires sufficient quality of the geometric models of the anatomy. Further, ligament simulation is time-consuming and highly dependent on the morphological data. Nevertheless, mere anatomical information of soft tissue attachments along forearm bones could already improve the preoperative pipeline. For example, better implant positions could be chosen, when taking into account the presence of soft tissue. Also, time during surgery could be saved and healing times be improved, by generating less invasive intraoperative interventions. Still, in future work we will focus on using the meshes in biomechanical simulations in the context of computer-assisted forearm surgery planning.


\begin{acknowledgements}
This work was supported by the Swiss National Science Foundation, under grant number 325230L\_163308, and the Austrian Science Fund FWF, under grant number I 2545-N31. 
\end{acknowledgements}

\bibliographystyle{spphys}   

\bibliography{References}   

\end{document}